\documentclass[global,final]{svjour}
\usepackage{epsfig}
\usepackage{amssymb,amsbsy,amsmath}
\usepackage{times}
\newcommand{\op}[1]{%
    \fontdimen12\textfont3=2pt\fontdimen12\scriptfont3=1.4pt%
    \!\null\mathop{\vphantom{#1}\smash{#1}}\limits_{\sim}\null\!}
\newcommand{\xref}[1]{\protect\ref{#1}}
\newcommand{\figref}[1]{Fig.~\protect\ref{#1}}
\newcommand{\fmref}[1]{(\protect\ref{#1})}
\def\ket#1{\, | \, {#1} \, \rangle}
\newcommand{\tr}{\mbox{tr}}

\newcommand{\Eins}
           {\op{\;\smash{\raisebox{-0.5ex}{$\!\!\stackrel{\!\mbox{1}
            \hspace{-0.4ex}\rule[0.0ex]{0.06ex}{1.60ex}}{ }$}}}}
\newtheorem{prop}{Proposition}

\journalname{Eur. Phys. J. B}
\begin{document}
\title{Independent magnon states on magnetic polytopes}
\author{J\"urgen Schnack \and Heinz-J\"urgen Schmidt
\inst{1} 
\and 
Johannes Richter \and J\"org Schulenburg
\inst{2}
}
\offprints{J\"urgen Schnack}          % Insert a name or remove this line
\institute{Universit\"at Osnabr\"uck, Fachbereich Physik,
Barbarastr. 7, D-49069 Osnabr\"uck, Germany
\and 
Institut f\"ur theoretische Physik,
Universit\"at Magdeburg,
P.O. Box 4120, D-39016 Magdeburg, Germany}
\date{Received: date / Revised version: date}
\maketitle
\begin{abstract}
For many spin systems with constant isotropic antiferromagnetic
next-neighbour Heisenberg coupling the minimal energies
$E_{min}(S)$ form a rotational band, i.~e. depend approximately
quadratically on the total spin quantum number $S$, a property which
is also known as Land\'{e} interval rule. However, we find that
for certain coupling topologies, including recently synthesised
icosidodecahedral structures this rule is violated for high
total spins. Instead the minimal energies are a linear function
of total spin.  This anomaly results in a corresponding jump of
the magnetisation curve which otherwise would be a regular
staircase.

\keywords{independent magnons -- magnetisation -- geometric
frustration}\\
\textbf{PACS} 75.50.Xx, 75.10.Jm, 75.40.Cx
\end{abstract}
%
%%%%%%%%%%%%%%%%%%%%%%%%%%%%%%%%%%%%%%%%%%%%%%%%%%%%%%%%%%%%%%%%%%%%%%%%
\section{Introduction}
\label{s-1}

It appears that for spin systems with constant isotropic
antiferromagnetic next-neighbour Heisenberg exchange the minimal
energy $E_{min}(S)$ for given total spin quantum number $S$ is
typically a strictly convex function of $S$. For many spin
topologies like rings, cubes, icosahedra etc. this function is
very close to a parabola \cite{ScL:PRB01}.  For certain systems
this behaviour has been explained with the help of the
underlying sublattice structure \cite{BLL:PRB94}.
Experimentally this property has been described as ``following
the Land\'{e} interval rule"
\cite{TDP:JACS94,LGC:PRB97A,LGC:PRB97B,ACC:ICA00}.  In the
classical limit, where the single-spin quantum number $s$ goes
to infinity, the function $E_{min}(S)$ is even an exact parabola
if the system possesses co-planar ground states \cite{ScL:CRB}.

However, we find that for certain coupling topologies, including
the cuboctahedron and the icosidodecadron \cite{polytopes}, this
rule is violated for high total spins.  More precisely, for the
icosidodecadron the last four points of the graph of $E_{min}$
versus $S$, i.~e.~the points with $S=S_{max}$ to $S=S_{max}-3$,
lie on a straight line
%--------------------------------------------------------
\begin{eqnarray}
\label{E-3}
E_{min}(S)
&=&
60 J s^2
-
6 J s (30 s - S)
\ .
\end{eqnarray}
%--------------------------------------------------------
An analogous statement holds for the last three points of the
corresponding graph for the cuboctahedron. These findings are
based on numerical calculations of the minimal energies for
several $s$ both for the icosidodecahedron as well as for the
cuboctahedron.  For both systems, additionally, we have a
rigorous proof of the high spin anomaly for the case of
$s=1/2$. This proof rests on an inequality which says that all
points of the graph of $E_{min}$ versus $S$ lie above or on the
line connecting the last two points (``bounding line"). The
proof can be easily applied to a wide class of spin systems,
e.g. to two-dimensional spin arrays.

The observed anomaly -- linear instead of parabolic dependence
-- results in a corresponding jump of the magnetisation curve
${\mathcal M}$ versus $B$. In contrast, for systems which obey
the Land\'{e} interval rule the magnetisation curve at very low
temperatures is a staircase with equal steps up to the highest
magnetisation.

The anomaly could indeed be observed in magnetisation
measurements of the so-called Keplerate structure
\{Mo$_{72}$Fe$_{30}$\} which is a recently synthezised magnetic
molecule where 30 Fe$^{3+}$ paramagnetic ions (spins $s=5/2$)
occupy the sites of a perfect icosidodecahedron and interact via
isotropic, next-neighbour antiferromagnetic exchange
\cite{MSS:ACIE99}.  Unfortunately, the magnetisation
measurements \cite{MLS:CPC,LSM:EPL} performed so far suffer from
too high temperatures which smear out the anomaly.

Nevertheless, it may be possible to observe truely giant
magnetisation jumps in certain two-dimensional spin systems
which possess a suitable coupling topology. In such systems the
magnetisation jump can be of the same order as the number of
spins, i.e. the jump remains finite -- or is macroscopic -- in
the thermodynamic limit $N\rightarrow\infty$.

The article is organized as follows. In section \xref{s-2} we
introduce basic definitions and explain how the results have
been obtained. In section \xref{s-3} the high spin anomaly is
discussed and proven for the case of $s=1/2$. We provide an
outlook in section  \xref{s-4}.

%%%%%%%%%%%%%%%%%%%%%%%%%%%%%%%%%%%%%%%%%%%%%%%%%%%%%%%%%%%%%%%%%%%%%%%%
\section{Definitions and Numerics}
\label{s-2}

\begin{figure}[ht!]
\begin{center}
\epsfig{file=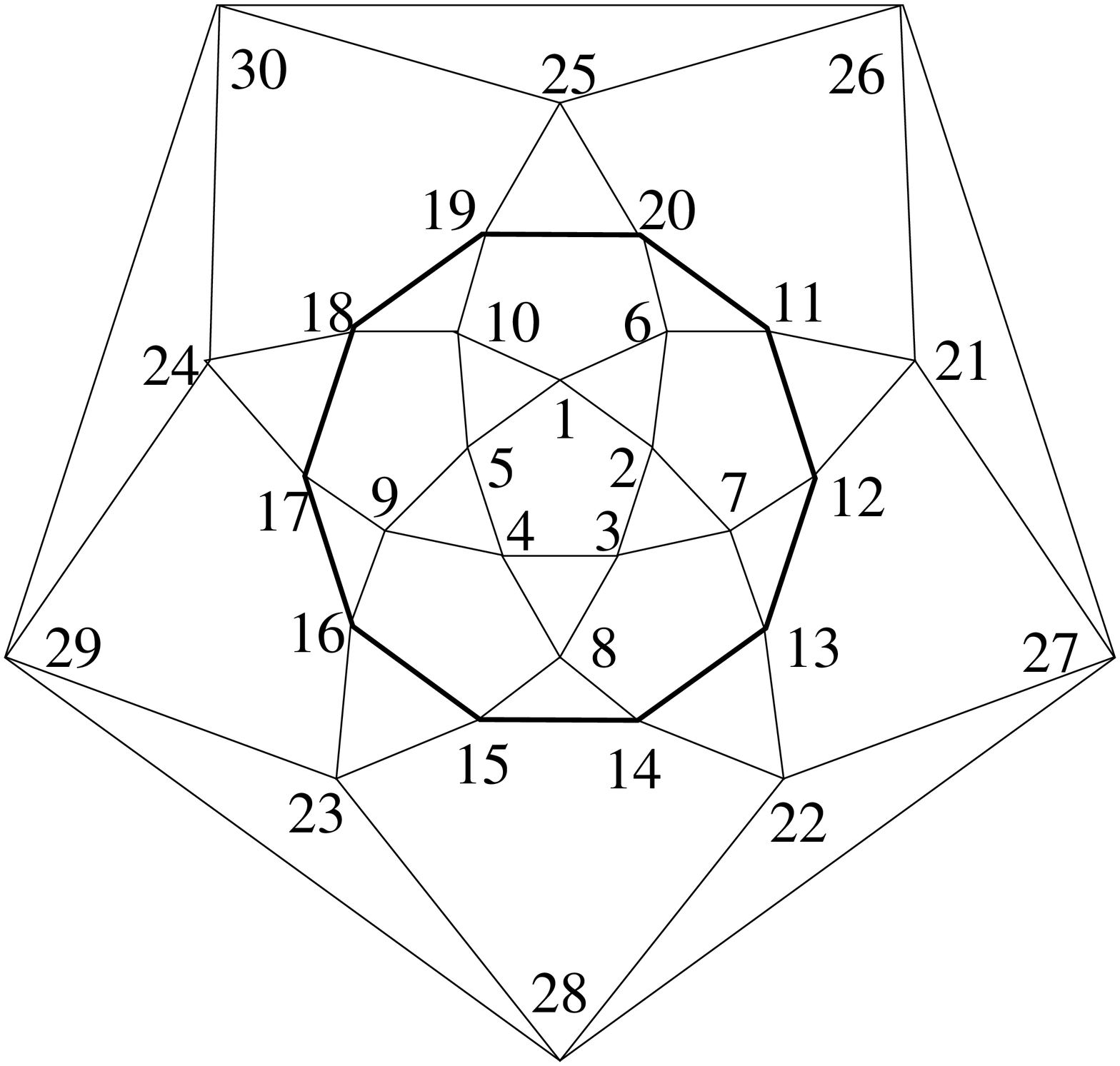,width=40mm}
$\quad$
\epsfig{file=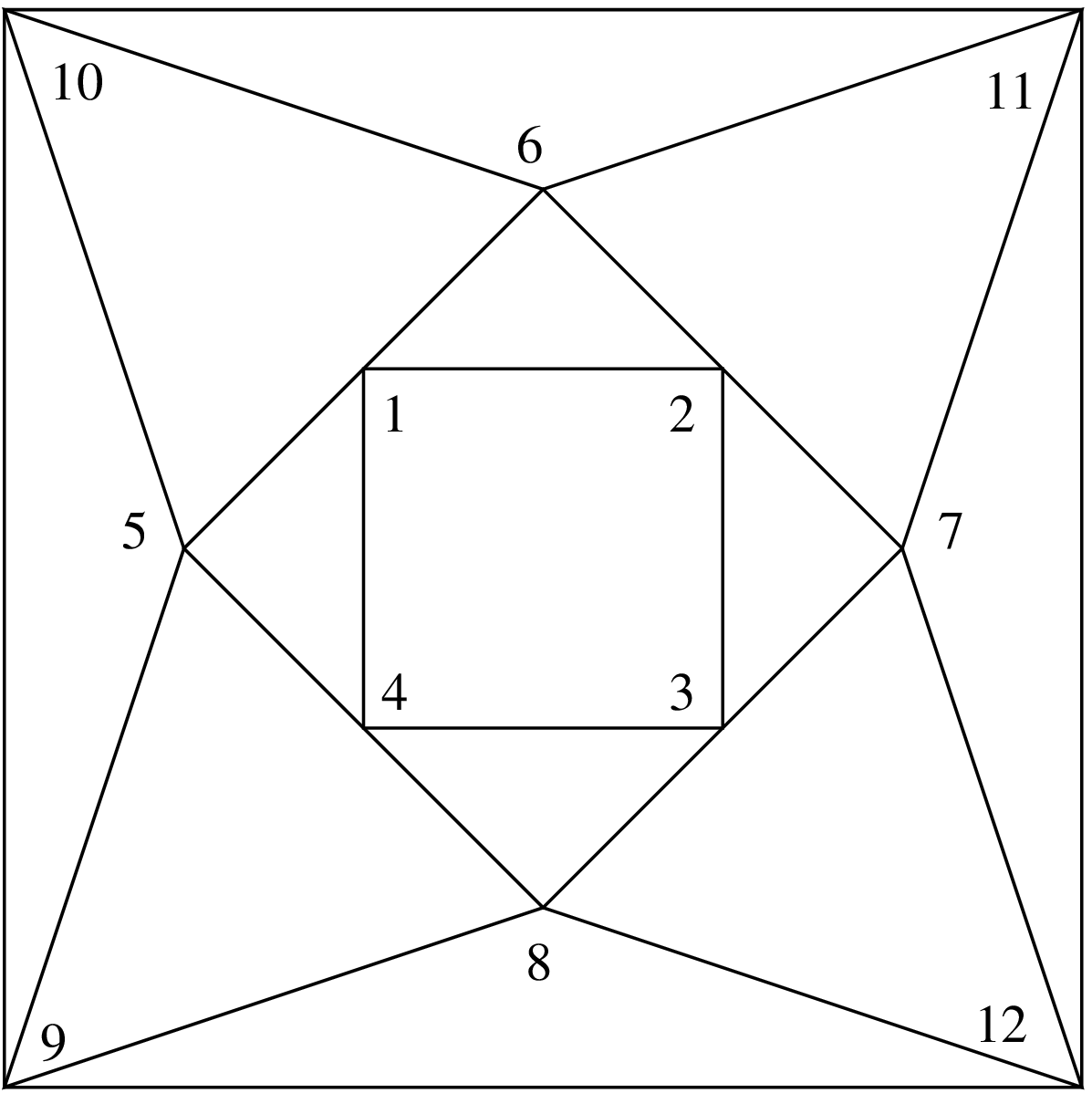,width=40mm}
\vspace*{1mm}
\caption[]{Planar projection of an icosidodecahedron (l.h.s.)
and a cuboctahedron (r.h.s.) \cite{polytopes}. Solid lines
denote couplings with a single exchange parameter $J$.}
\label{F-1}
\end{center}
\end{figure}

The Heisenberg Hamilton operator of the investigated spin
systems is 
%--------------------------------------------------------
\begin{eqnarray}
\label{E-1}
\op{H}
&=&
\frac{1}{2}
\sum_{(u,v)}\;
J_{u,v}\,
\op{\vec{s}}(u) \cdot \op{\vec{s}}(v)
+
g \mu_B B \op{S}_z
\\
&=&
\frac{J}{2}\,
\sum_{(u,v)\in\Gamma}\;
\op{\vec{s}}(u) \cdot \op{\vec{s}}(v)
+
g \mu_B B \op{S}_z
\ ,\quad
\op{S}_z = \sum_{u}\op{s}_z(u)
\ ,
\nonumber
\end{eqnarray}
%--------------------------------------------------------
where the exchange parameters $J_{u,v}$ are considered as the
components of a symmetric matrix $\mathbb{J}$, i.e. every bond
is taken into account twice.  In particular, we assume
$J_{u,v}\in\{ J,0 \}$ and $J>0$ which corresponds to
antiferromagnetic coupling. In equation (\ref{E-1}), $g$ is the
spectroscopic splitting factor and $\mu_B$ the Bohr
magneton. The vector operators $\op{\vec{s}}(u)$ are the spin
operators (in units of $\hbar$) of the individual $N$
paramagnetic ions with constant spin quantum number $s$.
Because the matrix $\mathbb{J}$ couples only next-neighbours
(see Fig.~\xref{F-1}) the second sum in
\eqref{E-1} runs over the set $\Gamma$ of all next-neighbour
pairs $(u,v)$ of spins of a single molecule at sites $u$ and
$v$. $\Gamma$ can be regarded as the set of ``edges" of the
corresponding undirected graph describing the coupling scheme of
the molecule. The ``vertices" of the graph correspond to the
spin sites $1,2,\dots,N$.  For each spin site $u$ let
$\Gamma(u)$ denote the set of neighbours of $u$. Throughout this
article we will assume that the number of neighbours per site is
constant, say $|\Gamma(u)|\equiv j$.  The ``distance" between
two spin sites $u$ and $v$ will be the minimal number of edges
connecting $u$ and $v$ (similar to the Manhattan distance).
%===================    figure   =================================
%===================    figure   =================================

As mentioned already in the introduction the anomaly was found
numerically. For this purpose the Hamilton matrix had to be
diagonalized. The total matrix is a huge object of dimension
$(2s+1)^N\times(2s+1)^N$ which must be block-diagonalized in
advance. Using that the Hamilton operator commutes with
$\op{S}_z$, the Ising product states which are a natural basis
can be grouped according to the quantum numbers $M$, thereby
dividing the Hilbert space into orthogonal subspaces ${\mathcal
H}(M)$. A further reduction of dimension is achieved if the
symmetries of the spin array are exploited. The
icosidodecahedron for instance shows a tenfold shift symmetry
leading to Hilbert subspaces ${\mathcal H}(M,k)$ with
$k=0,\dots,9$.  Within these subspaces a Lanczos procedure was
applied in order to obtain the respective minimal energies.

%%%%%%%%%%%%%%%%%%%%%%%%%%%%%%%%%%%%%%%%%%%%%%%%%%%%%%%%%%%%%%%%%%%%%%%%
\section{High spin anomaly}
\label{s-3}

%%%%%%%%%%%%%%%%%%%%%%%%%%%%%%%%%%%%%%%%%%%%%%%%%%%%%%%%%%%%%%%%%%%%%%%%%%%
\subsection{Observations}

\begin{figure}[ht!]
\begin{center}
\epsfig{file=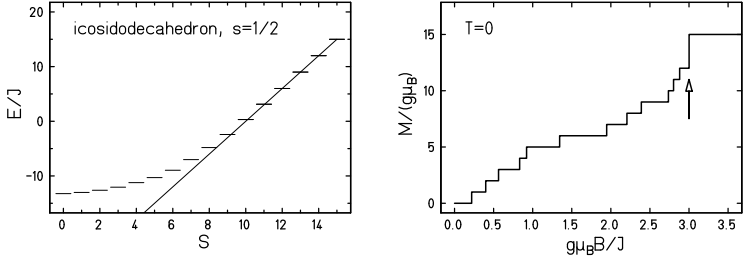,width=100mm}
\vspace*{1mm}
\caption[]{Icosidodecahedron: L.h.s. -- minimal energy levels
$E_{min}(S)$ as a function of total spin $S$. R.h.s. --
magnetisation curve at $T=0$.
}
\label{F-3}
\end{center}
\end{figure}

\begin{figure}[ht!]
\begin{center}
\epsfig{file=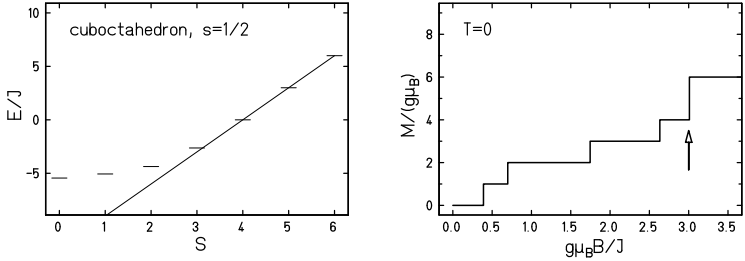,width=100mm}
\vspace*{1mm}
\caption[]{Cuboctahedron: L.h.s. -- minimal energy levels
$E_{min}(S)$ as a function of total spin $S$. R.h.s. --
magnetisation curve at $T=0$.}
\label{F-4}
\end{center}
\end{figure}

The resulting minimal energies $E_{min}(S)$ are shown by dashes
on the l.h.s. of \figref{F-3} for the isosidodecahedron and on
the l.h.s. of \figref{F-4} for the cuboctahedron. The straight
lines denote the bounding lines, which connect the highest four
levels in the case of the isosidodecahedron and the highest
three in the case of the cuboctahedron. At $T=0$ this behavior
leads to jumps of the magnetisation ${\mathcal M}$
%--------------------------------------------------------
\begin{eqnarray}
\label{E-2}
{\mathcal M}
&=&
-
\frac{1}{Z}\,
\tr\left\{g \mu_B \op{S}_z \mbox{e}^{-\beta\op{H}}\right\}
\ ,\quad
Z
=
\tr\left\{\mbox{e}^{-\beta\op{H}}\right\}
\ .
\end{eqnarray}
%--------------------------------------------------------
Due to the effect that the states lie exactly on the bounding
line in the graph of $E_{min}$ versus $S$ they ``take over" for
the new total ground state at the same value of the magnetic
field, therefore the magnetisation immediately jumps to the
highest value. The jumps are marked by arrows in the
magnetisation curves of the isosidodecahedron (r.h.s. of
\figref{F-3}) and the cubeoctahedron (r.h.s. of \figref{F-4}).
%===================    figure   =================================
%===================    figure   =================================
%===================    figure   =================================
%===================    figure   =================================

For systems which follow the Land\'{e} interval rule, i.e. where
$E_{min}(S)$ is a parabolic function of $S$, the corresponding
magnetisation curve would consist of equal steps.

%%%%%%%%%%%%%%%%%%%%%%%%%%%%%%%%%%%%%%%%%%%%%%%%%%%%%%%%%%%%%%%%%%%%%%%%%%%
\subsection{Idea of the proof}

A necessary condition for the anomaly is certainly that the
minimal energy in the one-magnon space is degenerate.
Therefore, localized one-magnon states can be constructed which
are also of minimal energy. When placing a second localized
one-magnon eigenstate on the spin array there will be a chance
that it does not interact with the first one if a large enough
separation can be achieved. This new two-magnon state is likely
the state of minimal energy in the two-magnon Hilbert space
because for antiferromagnetic interaction two-magnon bound
states do not exist (at least for $s=1/2$). This procedure can
be continued until no further independent magnon can be placed
on the spin array. In a sense the system behaves as if it
consists of non-interacting bosons which, up to a limiting
number, can condense into a single-particle ground state.

In more mathematical terms: In order to prove the high-spin
anomaly we first show an inequality which says that all points
$(S,E_{min}(S))$ lie above or on the line connecting the last
two points.  This inequality holds for $s=1/2$ and all systems
with constant antiferromagnetic exchange parameter and a
constant number of neighbours for each spin site. For specific
systems as those mentioned above what remains to be done is to
construct particular states which exactly assume the values of
$E_{min}$ corresponding to the points lying on the bounding
line, then these states are automatically states of minimal
energy.

Note that the high spin anomaly does not contradict the strict
convexity of the graph of $E_{min}$ versus $S$ in the classical
limit, since in the limit $s\rightarrow\infty$ the interval
where the anomaly occurs, e.~g.~$S=S_{max}-3$ to $S=S_{max}$,
becomes an infinitesimally small fraction of the total spin
range.

We set $J=1$ throughout this section.

%%%%%%%%%%%%%%%%%%%%%%%%%%%%%%%%%%%%%%%%%%%%%%%%%%%%%%%%%%%%%%%%%%%%%%%%%%%
\subsection{Bounding line for $s=1/2$}

Let ${\cal H}_a$ denote the eigenspace of $\op{S}_z$ with
eigenvalue $M=N/2-a,\quad a=0,1,\dots,N$. It has the dimension
%${\dim}({\cal H}_a)={N\choose a}$. 
${\dim}({\cal H}_a)=\begin{pmatrix}N\\a\end{pmatrix}$.  An
orthonormal basis of ${\cal H}_a$ is given by the product states
denoted by $|n_1,\dots,n_a\rangle$ with $1\le n_1 < n_2 <\dots
< n_a\le N$ where the $n_i$ denote the sites with flipped spin
$m=-1/2$. A state of this form will be called
\underline{isolated} iff $(n_i,n_j)\notin \Gamma$ for all $1\le
i<j \le a$.  In other words, the flipped sites of an isolated
state must not be neighbours according to the coupling
scheme. ${\cal H}_a^{iso}$ will denote the subspace of ${\cal
H}_a$ spanned by isolated states.

We will embed the Hilbert space of the spin system into some
sort of Fock space for magnons. More precisely, let ${\cal
B}_a({\cal H}_1)$ be the totally symmetric (i.~e.~bosonic)
subspace of $\bigotimes_{i=1,\dots,a} {\cal H}_1 $. If
$\op{A_1}: {\cal H}_1 \longrightarrow {\cal H}_1 $ is a linear
operator, ${\cal B}_a(\op{A_1})$ will denote the restriction of
$\op{A_1}\otimes\Eins\otimes\dots\otimes\Eins + \dots +
\Eins\otimes\dots\otimes\Eins\otimes \op{A_1}$ onto ${\cal
B}_a({\cal H}_1)$.  An orthonormal basis of ${\cal B}_a({\cal
H}_1)$ is given by the bosonic states
%--------------------------------------------------------
\begin{equation}\label{4.1}
\op{\cal S} \ket{n_1}\otimes\dots\otimes\ket{n_a},
\end{equation}
%--------------------------------------------------------
where $1\le n_1 < n_2 <\dots < n_a\le N$ and the
``symmetrisator" $\op{\cal S}$ denotes the sum over all $a!$
permutations of the product state divided by $\sqrt{a!}$.  The
linear extension of the map $\ket{n_1,\dots,n_a}\mapsto \op{\cal
S} \ket{n_1}\otimes\dots\otimes\ket{n_a}$ defines an isometric
embedding
%--------------------------------------------------------
\begin{equation}\label{4.2}
\op{\cal J}_a
:
{\cal H}_a
\longrightarrow
{\cal B}_a({\cal H}_1)
\ .
\end{equation}
%--------------------------------------------------------

Let $\op{H}_a$ denote the restriction of the Hamilton operator
(\ref{E-1}) (with zero magnetic field) onto ${\cal H}_a$ and
%--------------------------------------------------------
\begin{equation}\label{4.3}
\tilde{\op{H}_a} \equiv \op{\cal J}_a^\ast {\cal B}_a(\op{H}_1)\op{\cal J}_a.
\end{equation}
%--------------------------------------------------------

We will show the following
%%%%%%%%%%%%%%%%%%%%%%%%%%%%%%%%%%%%%%%%%%%%%%%%%%%%%%%%%%%%%%%%%%%%%%%%%%%
\begin{prop}\label{P1}
$\op{H}_a = \frac{1-a}{8}Nj + \tilde{\op{H}_a}$
if restricted to the subspace ${\cal H}_a^{iso}$.
\end{prop}
%%%%%%%%%%%%%%%%%%%%%%%%%%%%%%%%%%%%%%%%%%%%%%%%%%%%%%%%%%%%%%%%%%%%%%%%%%%
$j$ is the number of neighbours, which is assumed to be constant
for each spin site.

{\bf Proof}: Let $\ket{n_1,\dots,n_a}$ be an arbitrary isolated
basis state and split $\op{\vec{s}}(u)\cdot \op{\vec{s}}(v)$ into
%--------------------------------------------------------
\begin{equation}\label{4.4}
\op{\vec{s}}(u)\cdot \op{\vec{s}}(v) =
\op{s}_z(u)\op{s}_z(v)
+
\frac{1}{2}
(\op{s}^+(u)\op{s}^-(v)
+
\op{s}^-(u)\op{s}^+(v) )
\ ,
\end{equation}
%--------------------------------------------------------
analogously
$\op{H}_a=\op{H}_a^\prime+\op{H}_a^{\prime\prime}$
and
$\tilde{\op{H}_a}=\tilde{\op{H}_a}^\prime+\tilde{\op{H}_a}^{\prime\prime}$.
First, let us consider
%--------------------------------------------------------
\begin{eqnarray}\label{4.5}
&&
\op{H}_a^{\prime\prime} \ket{n_1,\dots,n_a}
=
\frac{1}{4}
\sum_{(u,v)\in\Gamma}
(\op{s}^+(u)\op{s}^-(v)
+
\op{s}^-(u)\op{s}^+(v) )
\ket{n_1,\dots,n_a} \\
&&=
\frac{1}{4}
\left(
\sum_{m_1\in\Gamma(n_1)}\mbox{Sort}\ket{m_1,\dots,n_a} 
+ \cdots +
\sum_{m_a\in\Gamma(n_a)}\mbox{Sort}\ket{n_1,\dots,m_a}
\right)
\nonumber
\ .
\end{eqnarray}
%--------------------------------------------------------
Here Sort denotes the procedure which re-arranges a list of
integers into its non-decreasing order. Note that further
summation constraints of the form $m_1\neq n_2$, $\dots$, $n_a$
etc.~would be superfluous since $\ket{n_1,\dots,n_a}$ was
assumed to be isolated.  Now consider
%--------------------------------------------------------
\begin{eqnarray}\label{4.6}
&&\tilde{\op{H}_a}^{\prime\prime} \ket{n_1,\dots,n_a}
=
\op{\cal J}_a^\ast
{\cal B}_a(\op{H}_1^{\prime\prime})
\op{\cal S}
\ket{n_1}\otimes\cdots\otimes\ket{n_a}\\
&&=
\frac{1}{4}
\op{\cal J}_a^\ast
\Bigg(
\sum_{m_1\in\Gamma(n_1)}
\op{\cal S}
\ket{m_1}\otimes\cdots\otimes\ket{n_a}
\nonumber
\\
&&
\qquad\qquad
+\cdots +
\sum_{m_a\in\Gamma(n_a)}
\op{\cal S}
\ket{n_1}\otimes\cdots\otimes\ket{m_a}
\Bigg)
\nonumber
\\
&&=
\frac{1}{4}
\left(
\sum_{m_1\in\Gamma(n_1)}\mbox{Sort}\ket{m_1,\dots,n_a} 
+ \cdots +
\sum_{m_a\in\Gamma(n_a)}\mbox{Sort}\ket{n_1,\dots,m_a}
\right)
\nonumber
\\
&&=
\op{H}_a^{\prime\prime} \ket{n_1,\dots,n_a}
\label{4.6-2}
\ .
\end{eqnarray}
%--------------------------------------------------------
Now we turn to $\op{H}_a^\prime$. Recall that there
is a total number of $L=\frac{Nj}{2}$ links between different
sites. For a given basis state
$\ket{n_1,\dots,n_a}$ we write
%--------------------------------------------------------
\begin{equation}\label{4.8}
L=L_{++}+ L_{+-}+L_{--},
\end{equation}
%--------------------------------------------------------
where  $L_{++}$ denotes the number of links between two $m=+1/2$-sites, etc.
Hence for isolated states $L_{--}=0$. Each basis state is an
eigenstate of $\op{H}_a^\prime$ with eigenvalue
%--------------------------------------------------------
\begin{equation}\label{4.9}
\frac{1}{4}(L_{++} - L_{+-}+L_{--})=\frac{1}{4}(L - 2 L_{+-}).
\end{equation}
%--------------------------------------------------------
For isolated states $L_{+-}=ja$, but in general $L_{+-}=ja-2 L_{--}$
since each $--$ link ``deletes" two $+-$ links of a corresponding
isolated state. Hence
%--------------------------------------------------------
\begin{eqnarray}\label{4.10}
\op{H}_a^\prime
\ket{n_1,\dots,n_a}
 &=&
\frac{1}{4}
\left(
\frac{Nj}{2} - 2ja +4 L_{--}
\right)
\ket{n_1,\dots,n_a}\\
&=&
\frac{1}{4}
\left(
\frac{Nj}{2} - 2ja
\right)
\ket{n_1,\dots,n_a}.
\label{4.10-2}
\end{eqnarray}
%--------------------------------------------------------
Similarly one can show that
%--------------------------------------------------------
\begin{eqnarray}\label{4.11}
\tilde{\op{H}_a}^\prime
\ket{n_1,\dots,n_a}
 &=&
\frac{a}{4}
\left(
\frac{Nj}{2} - 2j
\right)
\ket{n_1,\dots,n_a}.
\end{eqnarray}
%--------------------------------------------------------
From (\ref{4.11}), (\ref{4.10-2}), and (\ref{4.6-2}) the
proposition follows immediately.  \hfill $\blacksquare$

If we drop the condition that $\ket{n_1,\dots,n_a}$ is isolated we
ought to slightly modify our calculations. First, we would have to
introduce extra summation constraints of the form
$m_1\neq n_2,\dots,n_a$ etc.~in (\ref{4.5}) in order to be sure
that the resulting states lie in ${\cal H}_a$. Although
${\cal B}_a(\op{H}_1)$ in general will produce some unphysical
states with two magnons localized at the same site, these states
will be annihilated by $\op{\cal J}_a^\ast$. Hence again
$\op{H}_a^{\prime\prime}\ket{n_1,\dots,n_a} =
\tilde{\op{H}_a}^{\prime\prime}\ket{n_1,\dots,n_a}.$

For $\op{H}_a^{\prime}$ and $\tilde{\op{H}_a}^{\prime}$ the situation
is different. The foregoing arguments show that the
difference between the left and the right hand side of proposition
\ref{P1} is some operator with eigenstates
$\ket{n_1,\dots,n_a}$ and corresponding eigenvalues $L_{--}$
which are $\ge 0$. This shows that proposition \ref{P1} generalizes
to
%%%%%%%%%%%%%%%%%%%%%%%%%%%%%%%%%%%%%%%%%%%%%%%%%%%%%%%%%%%%%%%%%%%%%%%%%%%
\begin{prop}\label{P2}
$\op{H}_a \ge \frac{1-a}{8}Nj + \tilde{\op{H}_a}$
.
\end{prop}
%%%%%%%%%%%%%%%%%%%%%%%%%%%%%%%%%%%%%%%%%%%%%%%%%%%%%%%%%%%%%%%%%%%%%%%%%%%

Now let $E_a$ denote the smallest energy eigenvalue of $\op{H}_a$.
Note that $E_{min}(S=N/2-a)\ge E_a$, since the energy eigenvalues for
given total spin quantum number $S$ are assumed
within each subspace of magnetic quantum number $M=-S,\dots,S$.
We expect that $E_{min}(S=N/2-a) = E_a$ holds generally for the spin systems
under consideration, this has been proven only for so-called
bi-partite systems \cite{LSM:AP61,LiM:JMP62}, but numerically
shown to hold for much more systems \cite{RIV:JLTP99}.

Analogously we define $\tilde{E_a}$ for
$\tilde{\op{H}_a}$. Since for bosons the ground state energy is additive,
$a E_1$ will be the smallest energy eigenvalue of ${\cal B}(\op{H}_1)$.
We further conclude $\tilde{E_a}\ge a E_1$ since
$\tilde{E_a} = \langle \Phi | \tilde{\op{H}_a} | \Phi \rangle =
\langle \op{\cal J}_a \Phi | {\cal B}(\op{H}_1) | \op{\cal J}_a \Phi \rangle $
if $\tilde{\op{H}_a} \ket{\Phi} = \tilde{E_a} \ket{\Phi} $. Together
with proposition \ref{P1} this implies
%%%%%%%%%%%%%%%%%%%%%%%%%%%%%%%%%%%%%%%%%%%%%%%%%%%%%%%%%%%%%%%%%%%%%%%%%%%
\begin{prop}\label{P3}
$E_a \ge \frac{1-a}{8}Nj + a E_1$
.
\end{prop}
%%%%%%%%%%%%%%%%%%%%%%%%%%%%%%%%%%%%%%%%%%%%%%%%%%%%%%%%%%%%%%%%%%%%%%%%%%%

This inequality says that the minimal energies $E_a$, resp.~$E_{min}(S=N/2-a)$,
lie above or on the ``bounding line" $\ell(a)=\frac{1-a}{8}Nj + a E_1$.

%%%%%%%%%%%%%%%%%%%%%%%%%%%%%%%%%%%%%%%%%%%%%%%%%%%%%%%%%%%%%%%%%%%%%%%%%%%
\subsection{Ground states of independent magnons}
\label{sec-3.4}

According to what has been said above in order to rigorously
prove the high spin anomaly it suffices to construct states
which assume the energy values of the bounding line $\ell(a)$
for certain values of $a>1$. By the results of the previous
subsection it is clear that these energy values must be minimal
and the states must be eigenstates of $H_a$ in the case
$s=1/2$. Actually we conjecture that these states are also
minimal energy states of $H_a$ for arbitrary spin, which
conjecture is numerically supported for all cases where we have
calculated $E_a$, but we cannot prove it at the moment.
Nevertheless, we will assume an arbitrary spin $s$ in this
subsection.

We first consider the case of the icosidodecahedron. Let $a=1$.
Recall that a general state in ${\cal H}_a$ is of the form
$\sum_{n=1}^N c_n \ket{n}$, where $n$ denotes the spin
site where the magnetic quantum number is decreased by 1. The
eigenvalues of $H_1$ are of the form
%--------------------------------------------------------
\begin{equation}\label{5.1}
E_\alpha  = \frac{1}{2}Njs^2 +(j_\alpha - j)s ,
\end{equation}
%--------------------------------------------------------
where $j_\alpha , \alpha=1,\dots,N$ are the eigenvalues of the coupling
matrix $\mathbb{J}$. In our case, $N=30, j=4,$ and
the minimal eigenvalue $j_\alpha$ is $-2$,
hence
%--------------------------------------------------------
\begin{equation}\label{5.2}
E_1= 60 s^2-6s.
\end{equation}
%--------------------------------------------------------
The corresponding eigenspace of $H_a$ is ten-fold degenerate.
It is possible to find linear superpositions which are states of
minimal energy and have some intuitive geometric interpretation
as localized one-magnon states corresponding to even subrings of
the icosidodecahedron.  These states have alternating amplitudes
$c_n=\pm 1$ for sites $n$ of the subring and vanishing
amplitudes for the remaining sites. The smallest even subrings
generating such states are the ``8-loops" circumscribing two
adjacent pentagons, e.~g. $(1,2,3,4,9,17,18,10)$ according to
the numbering of sites in \figref{F-1} Other even subrings are
the ``equators" with 10 sites or the ``curly equators" with 12
sites which need not be further considered here.

Now let $a=2$. If a two-magnon ground state lies on the bounding
line $\ell(a)$, as it is suggested by numerical diagonalization,
we would have
%--------------------------------------------------------
\begin{equation}\label{5.3}
E_2=60 s^2-12 s.
\end{equation}
%--------------------------------------------------------
In fact, this energy is assumed by the following state: Consider
two 8-loops $L_1, L_2$ with a distance of $2$,
e.~g. $L_1=(1,2,3,4,9,17,18,10)$ and
\linebreak
$L_2=(12,13,22,28,29,30,26,21)$ according to \figref{F-1}.
$\epsilon_{n_1}, n_1\in L_1$ and $\delta_{n_2}, n_2\in L_2$
denote the amplitudes which define the one-magnon ground-states
described above. Then a two-magnon ground-state with the energy of
(\ref{5.3}) can be defined by
%--------------------------------------------------------
\begin{equation}\label{5.4}
\Phi_2
=
\sum_{n_1\in L_1, n_2\in L_2}
\epsilon_{n_1} \delta_{n_2} \ket{n_1,n_2}.
\end{equation}
%--------------------------------------------------------
Since this state lies entirely in ${\cal H}_2^{iso}$ it can be
considered as a ground-state of two non-interacting magnons.

Unfortunately, an analogous construction of three mutually
isolated one-magnon states is no longer possible for $a=3$.
Here we have to determine an appropriate state by numerical
diagonalization. One possible state of three independent magnons
is a state which is completely symmetric under the action of the
symmetry group of the icosidodecahedron, i.e. the icosahedral
group with reflections ${\cal Y}_h$ of order $120$. Hence it
will suffice to define this state by assigning an amplitude to
only one triple of sites within each orbit of the symmetry
group.  The other triples obtained by applying symmetry
operations $g\in{\cal Y}_h$ to each site will have, by
definition, the same aplitude.  The complete definition of this
state (without normalization) can be found in
Table~\xref{T-1}. The calculation of the corresponding energy
$E_2=60 s^2-18s$ can be done by a computer algebra
software. Also this state lies entirely within ${\cal
H}_3^{iso}$.  Thus we have obtained a rigorous proof of the
anomaly also for the case $a=3$ and $s=1/2$.
%--------------------------------------------------------
\begin{table}
\caption[]{Definition of an ${\cal Y}_h$-symmetric
three-magnon ground-state by assignment
of amplitudes to representative triple states.}\label{T-1}
\begin{tabular}{lrr}
$\ket{n_1, n_2, n_3}$ & Length of orbit & Amplitude \\
\hline
$\ket{1, 3, 14}$ & $60$ & $1$\\
$\ket{1, 3, 15}$ & $120$ & $-1$\\
$\ket{1, 3, 22}$ & $120$ & $-1$\\
$\ket{1, 3, 23}$ & $120$ & $1$\\
$\ket{1, 3, 28}$ & $120$ & $1$\\
$\ket{1, 3, 29}$ & $60$ & $-2$\\
$\ket{1, 7, 15}$ & $120$ & $1$\\
$\ket{1, 7, 18}$ & $60$ & $1$\\
$\ket{1, 7, 23}$ & $120$ & $-1$\\
$\ket{1, 7, 24}$ & $120$ & $-1$\\
$\ket{1, 7, 29}$ & $60$ & $2$\\
$\ket{1, 8, 21}$ & $120$ & $-1$\\
$\ket{1, 8, 25}$ & $30$ & $2$\\
$\ket{1, 8, 26}$ & $120$ & $-1$\\
$\ket{1, 8, 27}$ & $120$ & $2$\\
$\ket{1, 8, 28}$ & $30$ & $-2 $\\
$\ket{1, 13, 16}$ & $20$ & $-2$\\
$\ket{1, 13, 24}$ & $60$ & $-1$\\
$\ket{1, 13, 30}$ & $60 $ & $2$\\
$\ket{1, 14, 30}$ & $20$ & $-1$\\
\end{tabular}
\end{table}
%--------------------------------------------------------

The case of the cuboctahedron is largely analogous, up to the
fact that here we have only one point of anomaly for $a=2$. The
corresponding two-magnon ground state can be constructed by using
two separated 4-loops, e.~g.~$(1,2,3,4)$ and $(9,10,11,12)$ in
Fig.~\xref{F-1} (r.h.s.).

%%%%%%%%%%%%%%%%%%%%%%%%%%%%%%%%%%%%%%%%%%%%%%%%%%%%%%%%%%%%%%%%%%%%%%%%%%%
\subsection{Generalization to the XXZ-model}

The above proof holds also for the more general
Hamiltonian of the XXZ-model
%--------------------------------------------------------
\begin{eqnarray}
\label{E-A}
\op{H}
&=&
\frac{J}{2}\,
\sum_{(u,v)\in\Gamma}\;
\left\{
\Delta
\op{s}_z(u) \op{s}_z(v)
+
\op{s}_x(u) \op{s}_x(v)
+
\op{s}_y(u) \op{s}_y(v)
\right\}
\ ,
\end{eqnarray}
%--------------------------------------------------------
with $\Delta \ge 0$.  Since the total spin $S$ is no longer a
good quantum number, the minimal energies $E_{min}$ have to be
considered as a function of the total magnetic quantum number
$M$ instead. For the existence of the bounding line and the
corresponding magnetisation jump this aspect is irrelevant. The
only change in the proof is a multiplication of
$\op{H}_a^{\prime}$ and $\tilde{\op{H}_a}^{\prime}$ by $\Delta$,
which does not change the argumentation.  Also the construction
of eigenstates, as carried out in subsection \ref{sec-3.4}, is
not altered by the anisotropy $\Delta$ in \fmref{E-A}, since
these states are isolated.

%%%%%%%%%%%%%%%%%%%%%%%%%%%%%%%%%%%%%%%%%%%%%%%%%%%%%%%%%%%%%%%%%%%%%%%%%%%
\section{Outlook}
\label{s-4}

\begin{figure}[ht!]
\begin{center}
\epsfig{file=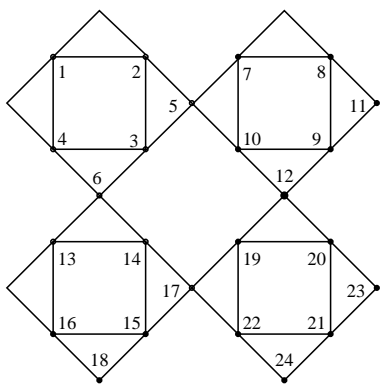,width=50mm}
\vspace*{1mm}
\caption[]{Fictitious two-dimensional spin array with periodic
boundary conditions. This array hosts at least as many
independent magnons as unit cells, i.e. $4\equiv N/6$.}
\label{F-5}
\end{center}
\end{figure}

The shown proof offers a method to create spin arrays which by
construction support a finite number of independent magnons. The
basic idea is to design a unit cell which can host a localized
one-magnon state, that is an eigenstate of the
Hamiltonian. Triangles play a key role in the construction of
such cells because they help to prevent localized magnons from
escaping.  The total spin array is then obtained by properly
linking serveral unit cells.  \figref{F-5} shows an example. The
unit cell is one quarter of the structure. It can host a single
magnon
%--------------------------------------------------------
\begin{equation}
\ket{\text{1 magnon}}
=
\frac{1}{2}
\left(
\ket{1}-\ket{2}+\ket{3}-\ket{4}
\right)
\ ,
\end{equation}
%--------------------------------------------------------
which is an eigenstate of the Hamiltonian with minimal energy in
the one-magnon space. One easily notices that in total four
localized independent magnons fit into the structure. In general
it might be possible that more independent magnons, like in the
case of the icosidodecahedron, can occupy the spin array. For
the example of \figref{F-5} this is not the case.
%===================    figure   =================================
%===================    figure   =================================

The latter example offers the perspective of observing truely giant
magnetisation jumps in two-dimensional spin systems. The number
of independent magnons which can be placed on the lattice is
proportional to the number of spins itself -- $N/6$ is the
example of \figref{F-5} -- and thus a macroscopic quantity. This
will be the subject of a forthcoming publication.

\begin{acknowledgement} 
J. Richter and J. Schulenburg thank the
Deutsche Forschungsgemeinschaft for support 
(project Ri 615/10-1).
\end{acknowledgement}

%%%%%%%%%%%%%%%%%%%%%%%%%%%%%%%%%%%%%%%%%%%%%%%%%%%%%%%%%%%%%%%%%%%%%%%%
%\bibliographystyle{/home/schnack/tex/bibtex/fmdplain.bst}
%\bibliography{/home/schnack/tex/bibtex/js.bib}

\end{document}